\documentclass{PoS}

\title{Dispersive analysis tools for $\pi\pi$ and $\pi K$ scattering}

\ShortTitle{Dispersive $\pi\pi$ and $\pi K$ scattering tools}

\author{\speaker{J.R. Pel\'aez}\\
        Departamento de F\'{\i}sica Te\'orica-UPARCOS. Facultad de CC. F\'{\i}sicas.\\
        Universidad Complutense. 28040 Madrid- SPAIN\\
        E-mail: \email{jrpelaez@fis.ucm.es}}

\author{A. Rodas\\
        Departamento de F\'{\i}sica Te\'orica-UPARCOS. Facultad de CC. F\'{\i}sicas.\\
        Universidad Complutense. 28040 Madrid- SPAIN\\
        E-mail: \email{arodas@fis.ucm.es}}

\abstract{In this talk we discuss the relevance of dispersive methods
and their different applications to obtain simple but rigorous parameterizations
of pion-pion and kaon-pion scattering. We show as an example
our latest dispersive analysis of $\pi K$ scattering, where we provide
simple parameterizations of data satisfying Forward Dispersion Relations up to 1.6 GeV.}

\FullConference{XVII International Conference on Hadron Spectroscopy and Structure - Hadron2017\\
		25-29 September, 2017\\
		University of Salamanca, Salamanca, Spain}

\begin{document}

Pions and kaons are the
pseudo-Goldstone bosons of the spontaneous
chiral symmetry breaking of QCD. This makes them the lightest non-strange and strange hadrons, 
respectively, and thus they are present in the final products of almost all hadronic interactions,
so that a good knowledge of their rescattering becomes essential. In addition, by studying their 
interactions we are testing the QCD chiral symmetry breaking pattern. Moreover, many of the 
lightest mesons 
appear as resonances in $\pi\pi$ or $\pi K$  scattering. In particular, this is the case of some of the most 
controversial states, like the lightest scalars, whose composition is likely different
from the ordinary quark-antiquark mesons of the simple quark model.

Experimental information on $\pi\pi$ and $\pi K$ scattering can only be obtained indirectly.
The main data sets come from meson-nucleon to meson-meson-nucleon experiments performed in the 70's and 80's
 (see \cite{Pelaez:2015qba} for references). By assuming that these processes are dominated by one-pion exchange, one can extract meson-pion to meson-pion
scattering as a subprocess. Unfortunately, the extraction of the meson-pion scattering subprocesses 
is plagued with systematic uncertainties due to the on-shell extrapolation of the exchanged pion, other possible exchange mechanisms, rescattering models, ambiguities, etc... As a consequence there are several conflicting sets of data both for $\pi\pi$ and $\pi K$ scattering in the literature. In the case of $\pi\pi$ scattering, very precise data has been obtained from $K\rightarrow\pi\pi \ell \nu_\ell$, but is is limited to energies below the kaon mass and a particular combination of isospins.

Concerning  resonances, apart from the difficulties arising from conflicting data sets,
they have been frequently determined in terms of simple models. 
Unfortunately, the most interesting resonances are rather wide or lie near thresholds, 
where simple models become unreliable. 

In this context, dispersive methods have become an essential tool for
model-independent analyses. 
Let us recall that the scattering amplitude $T(s,t,u)$ actually depends on two of the $s,t,u$ usual Mandelstam variables, 
since $s+t+u=\sum_i m_i^2$ and $m_i$ are the masses of the two initial and two final mesons.
Dispersion relations are based on the analyticity of the amplitudes in the complex $s,t$ plane, which is the mathematical realization of causality.
For most purposes, one considers the amplitude as a function of one variable either by fixing the other or by integrating it out in partial waves. In both cases, and since there are no bound states, the amplitudes are analytic in the complex $s$ plane except for a ``physical'' or ``right'' cut from threshold to infinity and a ``left'' cut that accounts for thresholds in the crossed channels and extends to minus infinity. In the case of unequal masses, partial waves develop an additional  circular cut.  
Then Cauchy Theorem relates 
the value of the amplitude at a given energy in terms of an integral of the amplitude over 
the cuts. The evaluation over the ``unphysical'' cuts is usually the most complicated part.
If the aim is not high precision, one can then think about approximating these unphysical 
cut contributions and the subtraction constants with Chiral Perturbation Theory,
which is the QCD low-energy expansion, see for instance \cite{Zheng:2003rw}.
This idea is also the basis
of unitarized Chiral Perturbation Theory, although in this case the dispersion relation is written for the inverse of the partial wave, since its imaginary part is known exactly from elastic unitarity
on the right cut. The results are fairly good to describe data and since they are based on analyticity allow for a sound analytic continuation to the complex plane and a fairly good determination
of resonances in the elastic region. This is known as the elastic Inverse Amplitude Method (IAM) \cite{IAM}.
The treatment of the coupled channel case with dispersion relations is not straightforward and usually made through a naive extension of the IAM or the N/D method, although the left cut is often neglected or very crudely approximated (see \cite{Pelaez:2015qba} for details). 

But if one is aiming for precision and a rigorous treatment of the left cut using data,
two main kinds of dispersion relations have been used in the literature:



{\bf 1) Fixed-variable dispersion relations}. One variable is fixed once the other is known. The most common are fixed-$t$ dispersion relations, but hyperbolic dispersion relations, where two variables are forced to lie in an hyperbola designed to maximize the applicability domain are also used, particularly for $\pi K$ (or $\pi N$, $\pi\pi\rightarrow\bar{K}K$ and $\gamma\gamma\rightarrow \pi\pi$) scattering. 
Of special relevance are Forward Dispersion Relations (FDR), with fixed $t=0$, 
since the left cut is easily rewritten in terms of the physical region were the forward amplitude is related by the optical theorem to the total cross section, 
for which, at high energies, there is usually more available data.

{\bf 2) Partial-wave dispersion relations}. Here one integrates one variable by means of the partial-wave expansion.  In practice, this limits their applicability to $\hat s\sim 1$GeV$^2$, i.e. little more than the elastic regime for the direct channel, and to $\hat t\sim 2 \hat s$ for the crossed channel. These relations are best suited to calculate resonance poles, which is the 
rigorous way to determine the existence of resonances and their parameters. 
In the elastic regime, these poles lie in the second Riemann sheet of the complex plane and dispersion relations are formulated on the first. However, the second sheet is easily accessible for partial waves due to their algebraic unitarity condition in the real axis, once the first sheet is known.  When crossing symmetry is used to rewrite the contributions from crossed channels (left cut) in terms of the direct channel physical region,  one gets a system of coupled system of infinite integral relations. For $\pi\pi$ scattering these are known as Roy equations \cite{Roy:1971tc}, whereas for scattering of particles with different masses these are customarily called Roy-Steiner equations \cite{Steiner:1971ms}. In practice the system is solved for the lowest partial waves at low energies considering the other waves and the high energy contributions
 as input.

In both cases, the high energy region contribution can be suppressed to make the integrals convergent 
or just because it is usually less known,
with powers of the energy in the integrand denominator (subtractions), 
but then the price to pay is that the amplitude is determined up to a polynomial on the variable under consideration, whose coefficients
are called subtraction constants.

Dispersion relations can be used mainly for four different purposes:
\begin{enumerate}
\item To check the consistency of data and possibly to discard some inconsistent data sets.

\item To obtain constrained parameterizations of data, consistent with dispersion relations.

\item To calculate the amplitude in a certain region as a solution of the dispersion relations, using as input the rest of the amplitude in other regions or in other channels.

\item To obtain rigorous analytic continuations of the amplitudes to the first Riemann sheet.
In the case of partial-wave dispersion relations, unitarity 
allows access to the second sheet and to a rigorous determination of the existence and parameters of 
resonances in the elastic regime.

\end{enumerate}

Let us briefly review these recent uses for $\pi\pi$ and $\pi K$ scattering.

\section{Dispersive analyses of $\pi\pi$ scattering}

Dispersive results are most relevant for the isoscalar $S-$wave since it is there that data show more conflicts between different experiments and where the controversial light scalar resonances appear  (see \cite{Pelaez:2015qba} for a review).
Apart from the IAM or N/D dispersive methods, for this process
there have been two main dispersive approaches aiming for precision:

1) To {\bf solve} Roy equations for $S-$ and $P-$waves in the low energy region 
below a matching point $s_m$,
using data parameterizations as input for higher waves and the for the $S-$ and $P-$waves above $s_m$
\cite{Ananthanarayan:2000ht,Colangelo:2001df}. In practice, the original twice-subtracted Roy equations were used, thus requiring information on the two $S-$wave scattering lengths. 
The results are in fairly good agreement with some of the existing sets of $\pi\pi$ scattering data.
When Chiral Perturbation Theory is used to constraint these
parameters, the precision is highly improved \cite{Colangelo:2001df}, 
leading also to a remarkably precise
prediction for the then controversial $f_0(500)$ or $\sigma$-meson \cite{Caprini:2005zr}, 
consistent with previous determinations with the IAM \cite{IAM}.
The matching point $s_m=(0.8 \rm GeV)^2$ of \cite{Ananthanarayan:2000ht,Colangelo:2001df}
was extended to the $K\bar{K}$ threshold for the scalar-isoscalar wave
in \cite{Moussallam:2011zg}, using for the rest the same input as in  \cite{Colangelo:2001df}.
The resulting $f_0(500)$ pole is very similar to that in \cite{Caprini:2005zr}
but this extension allowed for a rigorous determination of the $f_0(980)$ pole and the inelasticity in that region.

2) To use Dispersion Relations both {\bf as checks and constraints} for {\it fits to data}. 
This analysis was carried out by the Madrid-Krakow group.
On a first step it was shown that many data sets in the literature severely violated
dispersion relations \cite{Pelaez:2004vs}. Then the data sets close 
to satisfying these relations were refitted using dispersion relations as constraints.
This was done in a series of works \cite{Pelaez:2004vs,Kaminski:2006yv, GarciaMartin:2011cn} refining the fits and imposing three FDRs covering an isospin basis, 
the three Roy equations with two subtractions for $S-$ and $P-$waves, as well 
as Roy equations but with one-subtraction (called GKPY equations \cite{GarciaMartin:2011cn})
for the same three waves. FDRs were applied up to 1.4 GeV and Roy and GKPY equations
 up to 1.1 GeV. 
The latter produce smaller uncertainties in the resonance region than standard Roy equations, which are better at threshold. The result of this process is a simple set of partial-wave amplitudes that satisfy these nine dispersion relations (and some other sum rules), while still describing the data \cite{GarciaMartin:2011cn}. By reaching the region between 1 and 1.1 GeV it was possible to solve a longstanding data conflict between data on the inelasticity in that region, essential for the determination of the $f_0(980)$ resonance parameters.
In a later work \cite{GarciaMartin:2011jx}, Roy and GKPY equations were used to determine the pole of the $f_0(500)$, very consistent with the determinations from solutions of Roy eqs., as well as those for  the $f_0(980)$ and $\rho(770)$ mesons. 

Mainly as a consequence of these works, the uncertainties of the $\sigma$ meson parameters that were
listed in the Review of Particle Properties (RPP)
for years with a mass of 400 to 1200 MeV and a with of 500 to 100 MeV, were
reduced to a pole mass $M\simeq 400-550\,$MeV and a pole width of $\Gamma\simeq400-700\,$MeV in the RPP 2012 edition \cite{RPP2012}, changing the name to $f_0(500)$. Nevertheless in the very RPP it was argued that the ``most advanced dispersive analyses'' \cite{Ananthanarayan:2000ht,Caprini:2005zr,Moussallam:2011zg,GarciaMartin:2011jx}
produced much smaller uncertainties. Taking into account systematic uncertainties
this has been recently estimated as $M=449^{+22}_{-16}\,$MeV and $\Gamma=550\pm24\,$MeV
\cite{Pelaez:2015qba}. The central values and uncertainties of the $f_0(980)$ were also changed
in the RPP 2012 edition due to the dispersive evaluations in \cite{Moussallam:2011zg,GarciaMartin:2011jx}.

\section{Dispersive analyses of $\pi K$ scattering}

In this case, the crossed channel is a different reaction $\pi\pi\rightarrow \bar{K}K$.
When using crossing symmetry to rewrite the unphysical contributions into the physical
cut of $\pi K$ partial waves, one needs the Roy-Steiner formalism. 
Note that   
in order to reach the controversial $K^*_0(800)$ or $\kappa-$meson pole, whose situation in the RPP is similar to that of the $\sigma$ meson before 2012, the Roy-Steiner formalism should be based on
hyperbolic dispersion relations for a larger applicability region.
Actually, the best determination so far of the $K^*_0(800)$ meson comes from a {\it solution} 
of Roy-Steiner equations for $\pi K$ scattering in the elastic region \cite{Buettiker:2003pp} derived from fixed-$t$ dispersion relations and 
a posteriori analytic extension based on Roy-Steiner equations from hyperbolic relations 
\cite{DescotesGenon:2006uk}. Nevertheless the RPP still lists the $K^*_0(800)$ under ``needs confirmation'', and we have been encouraged to apply to $\pi K$ a similar approach to that of the Madrid-Krakow group for $\pi\pi$ and the $f_0(500)$ meson.

Thus, since Roy-Steiner equations are limited to $\sim$1.1 GeV, we have started by performing
an FDR analysis of $\pi K$ data \cite{Pelaez:2016tgi}. We have first shown that unconstrained
fits to data (UFD) do not satisfy well the FDRs, see the top panels in Fig.1, where for agreement the ``input'' line should fall
within the uncertainties of the ``Dispersive'' one.
Next we have obtained constrained fits to data (CFD) consistent with FDRs
up to 1.6 GeV, see the bottom panels in Fig.1.

In Fig.2 we show the comparison of the UDF with the CDF for the S-wave, 
which is the most interesting one. The change is not very large, except at high energies, where 
it seems to prefer one data set over the other (see \cite{Pelaez:2016tgi} for details and references).

\begin{figure}
\centering
\includegraphics[scale=0.28]{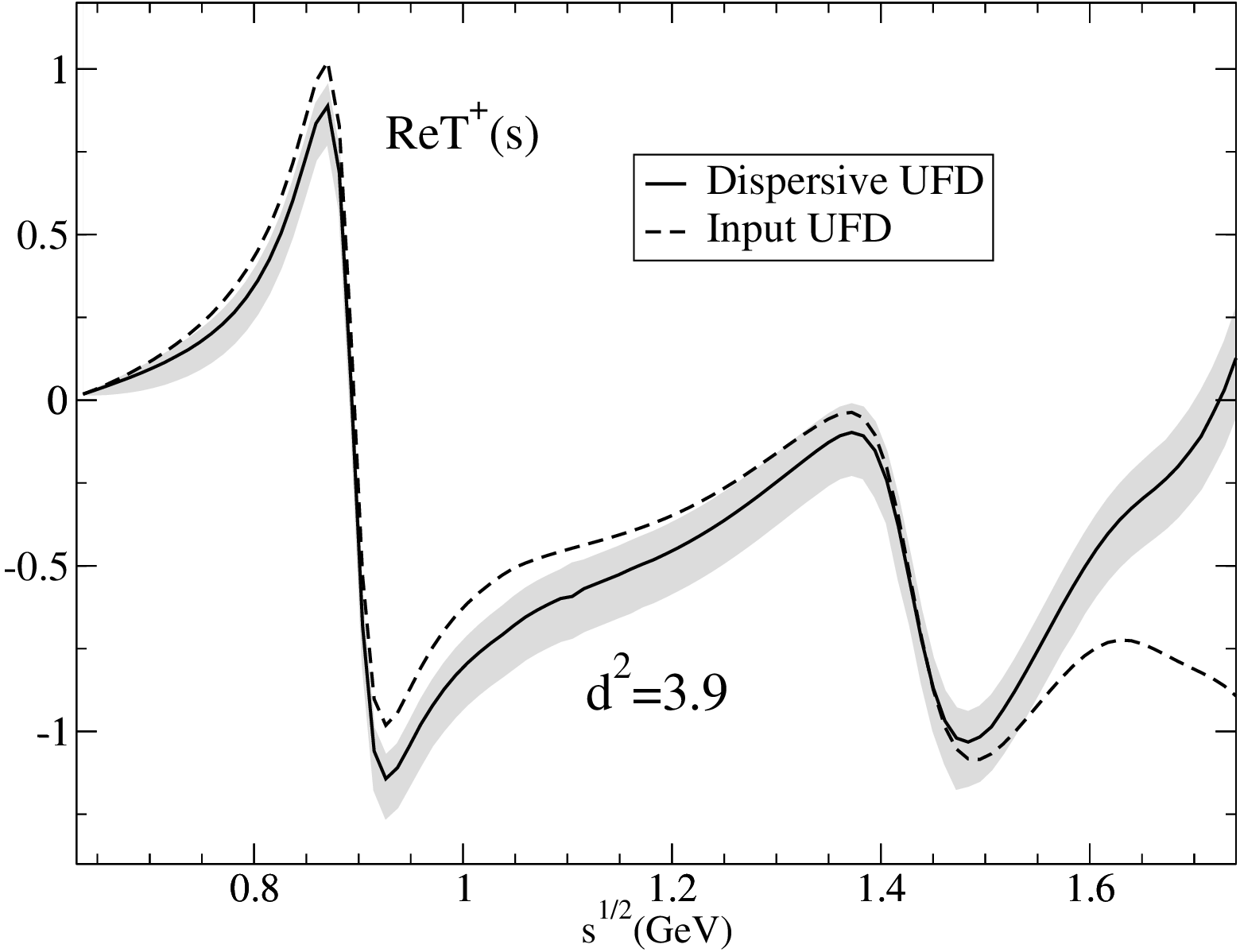}\hspace*{.3cm}
\includegraphics[scale=0.28]{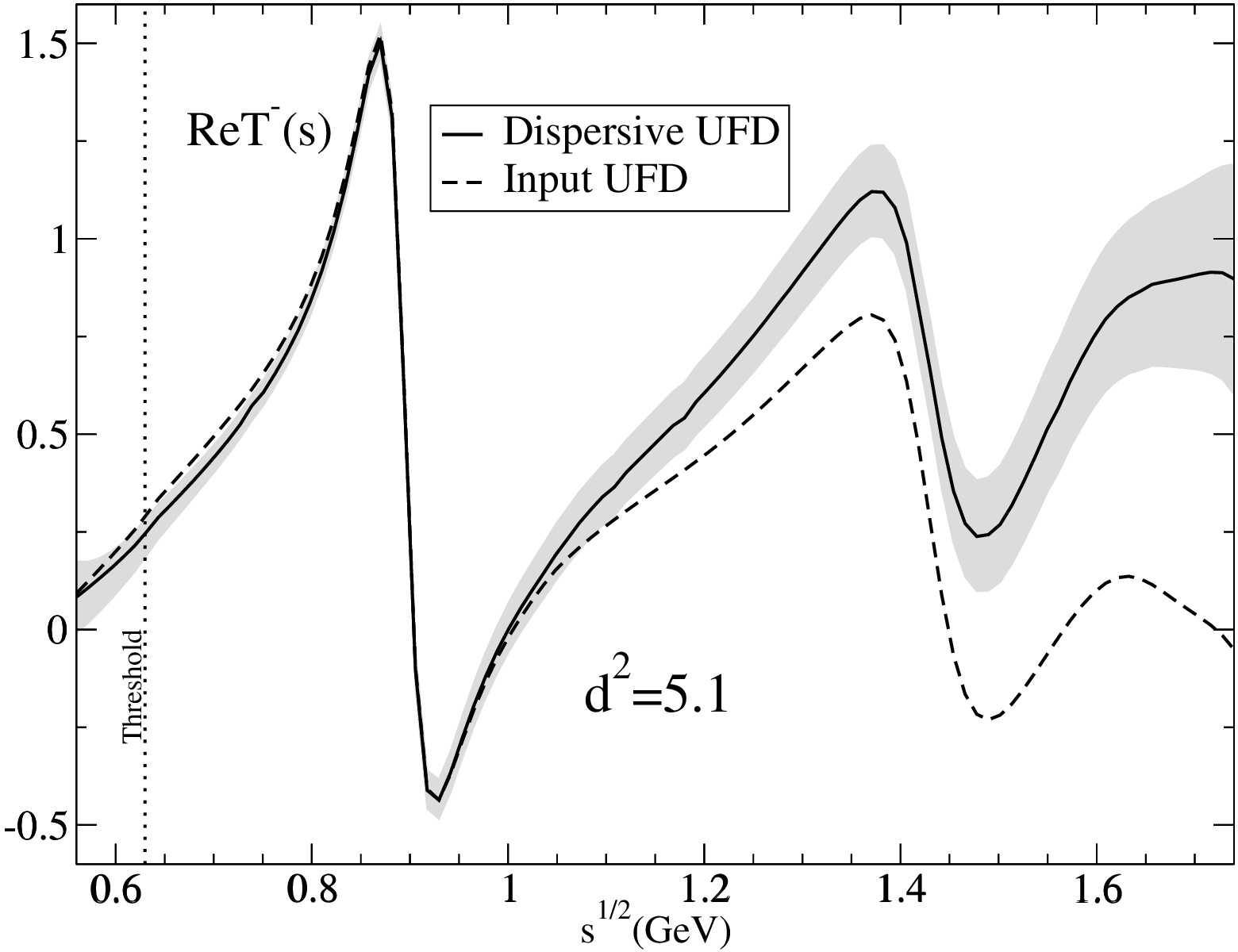}
\includegraphics[scale=0.28]{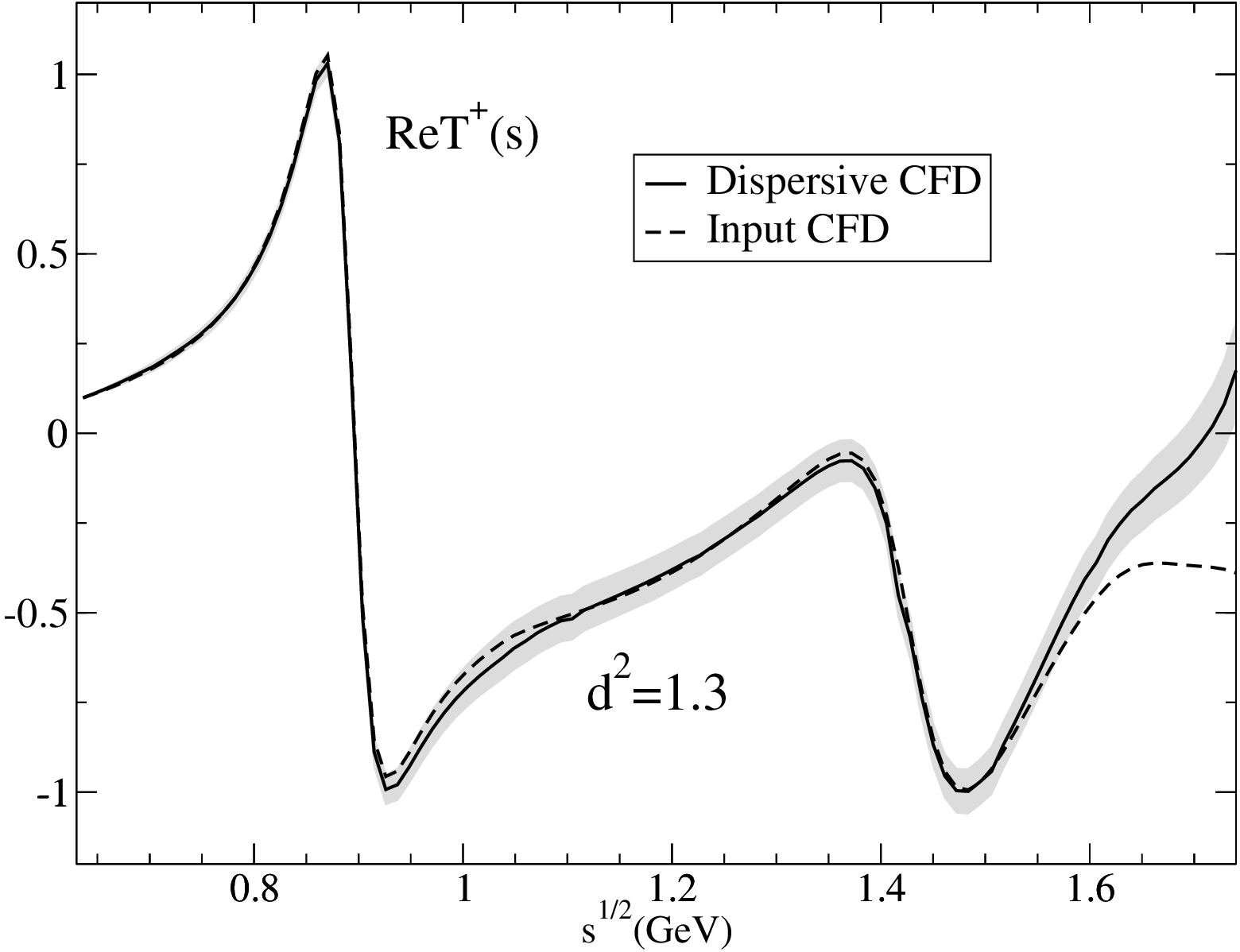}\hspace*{.3cm}
\includegraphics[scale=0.28]{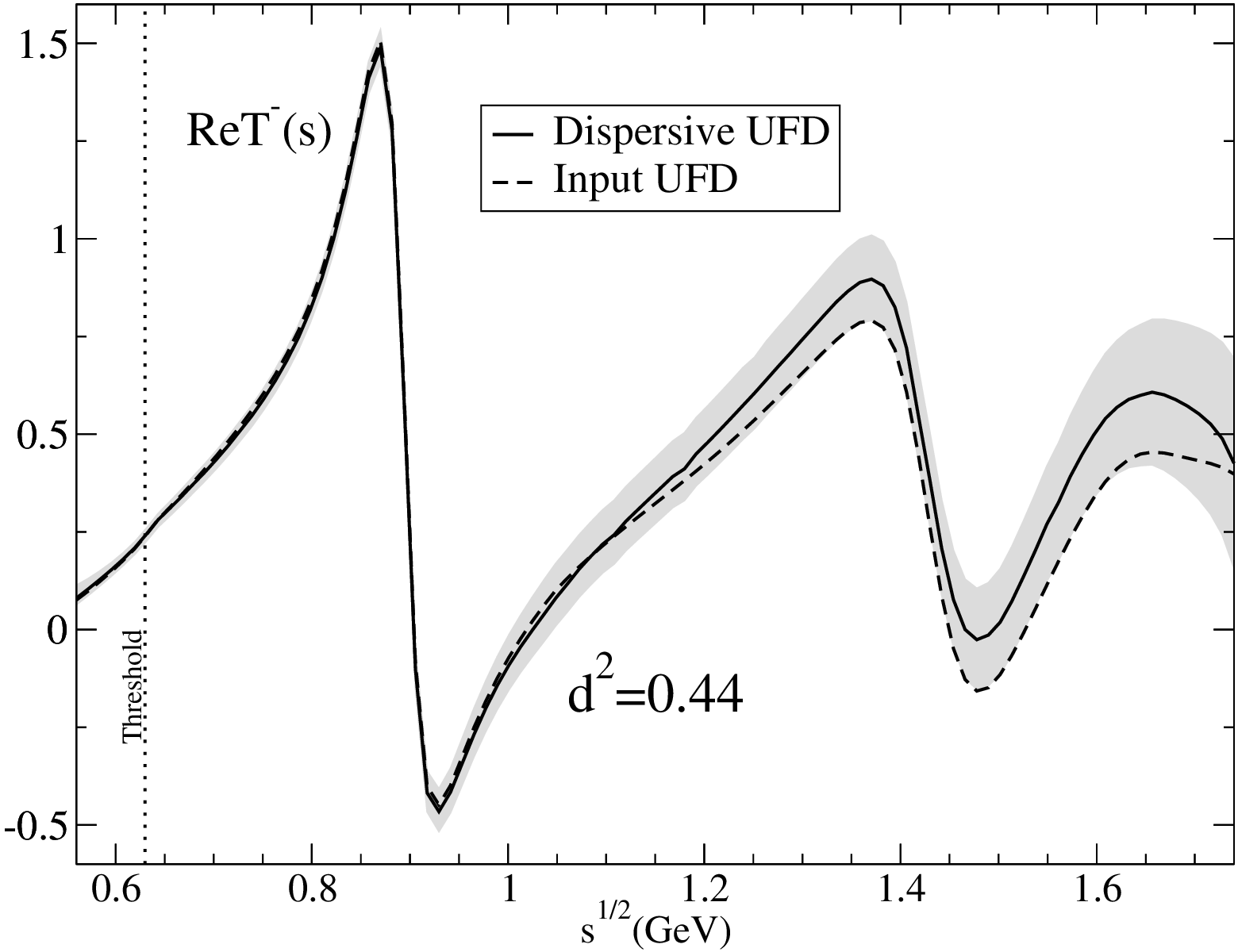}
\caption{\rm \label{fig:FDRUFD} 
Comparison between the input
and the dispersive result for the $T^+$ (left) and $T^-$(right) FDRs when using
the UFD (top) or the CFD (bottom). The CFD set is consistent within errors up to 1.6 GeV. Figures taken from  \cite{Pelaez:2016tgi}.
}
\end{figure}

\begin{figure}
\centering
\includegraphics[scale=0.73]{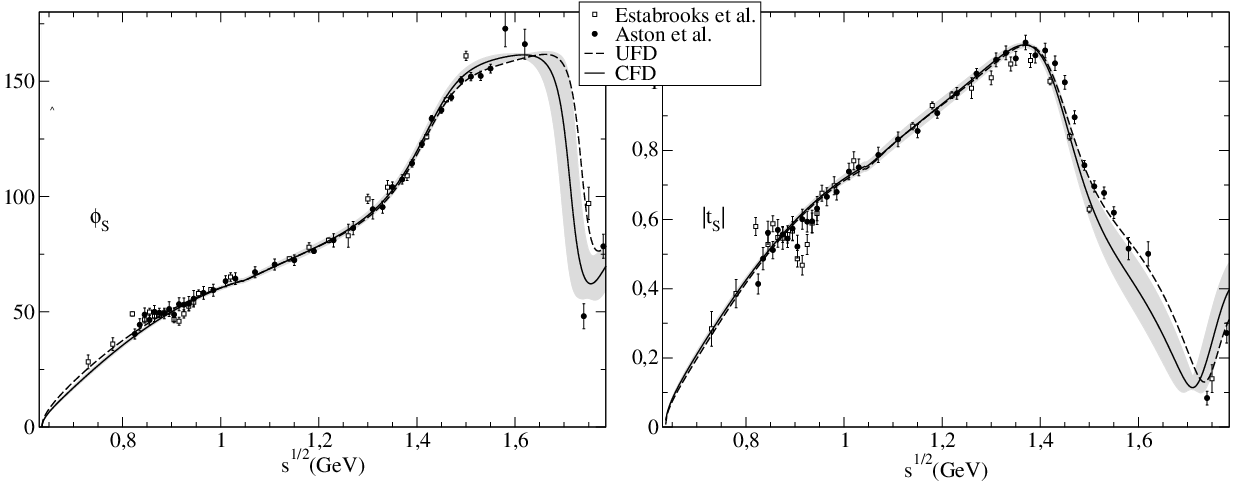}\hspace*{.3cm}
\caption{\rm \label{fig:FDRUFD} 
The unconstrained (UFD) versus constrained (CFD+error band) fits to isoscalar-scalar data.
Figures taken from  \cite{Pelaez:2016tgi}.
}
\end{figure}

Therefore, we have obtained \cite{Pelaez:2016tgi} a description of data which is simultaneously consistent with FDRs up to 1.6 GeV, in terms of simple parameterizations that are easy to implement for further phenomenological studies. In particular, apart from the interest on their own, they can be now used as input for further studies of $\pi K$ and $\pi\pi \rightarrow \bar{K}K$ scattering and the confirmation of the $K_0^*(800)$ resonance.

Concerning resonances, our fixed-$t$ dispersion relations can be analytically continued to the first Riemann sheet of the complex plane, but in order to get to the second sheet with dispersion relations
in search for resonance poles, we would need partial-wave dispersion relations (in preparation).
In the meantime, we can use in the elastic region our parameterization, which is
 nothing but a conformal expansion
with remarkably good analytic properties
up to the $\kappa$-pole, which we have obtained and is fairly consistent with the dispersive \cite{DescotesGenon:2006uk} results based on partial waves. 
For resonances in the inelastic region, our parameterizations are piecewise functions 
that cannot be naively continued to the complex plane. However, there are methods \cite{Masjuan:2014psa} based on sequences of Pad\'es that can be used to obtain a stable determination of 
the position  resonance poles, starting from the value of 
our CFD partial waves in the real axis provided in \cite{Pelaez:2016tgi}. This method allows for a 
determination of the existence and parameters of the $K_0^*(800)$, $K_0^*(1430)$, $K_1^*(892)$, $K_1^*(1410)$, $K_2^*(1430)$ and  $K_3^*(1780)$ resonances (see \cite{Pelaez:2016klv} and A. Rodas talk in this Conference).

\section{Summary}
We have reviewed the relevance of dispersion theory to discard conflicting 
$\pi\pi$ or $\pi K$ scattering data
and to obtain consistent parameterizations of scattering amplitudes, showing some recent uses of relevance for the determination of parameters of controversial resonances. We have summarized our latest work on a dispersive analysis of $\pi K$ scattering up to 1.6 GeV.
\section{Acknowledgments}
Work supported by the Spanish Projects  FPA2014-53375-C2-2,
FPA2016-75654-C2-2-P.
A. Rodas acknowledges 
the Universidad Complutense for a doctoral fellowship.

\end{document}